\documentclass[aps,pra,twocolumn,showpacs,superscriptaddress,footinbib,groupedaddress,longbibliography]{revtex4-1}
\usepackage{graphicx}  % needed for figures
\usepackage{dcolumn}   % needed for some tables
\usepackage{bm}        % for math
\usepackage{amssymb}   % for math
\usepackage{gensymb}   % for degree symbol
\usepackage{amsmath}
\usepackage{float}
\usepackage{xcolor}
\usepackage[english]{babel}
\usepackage{color}
\usepackage{ulem}

\usepackage[colorlinks=true,citecolor=black,urlcolor=blue,linkcolor=black]{hyperref} % choice of color for hyperref links

\newcommand{\vkeff}{\vec{k}_{\text{eff}}}
\newcommand{\keff}{k_{\text{eff}}}

\newcommand{\overbar}[1]{\mkern 1.5mu\overline{\mkern-1.5mu#1\mkern-1.5mu}\mkern 1.5mu}

\begin{document}

%\title{Pure rate-gyroscope based on cold-atom interferometry with varying momentum transfers}
%\title{Multi-loop atom interferometry with imbalanced momentum transfer}
%\title{Multi-loop atom interferometry with variable momentum transfers}
%\title{Multi-loop matter-wave interferometry with tilted mirrors}
%\title{Multi-loop matter-wave interferometry with tailored mirrors}
\title{Tailoring multi-loop atom interferometers with adjustable momentum transfer}

%\author{GYRO authors}
\author{L. A. Sidorenkov}
\email{leonid.sidorenkov@gmail.com}
\author{R. Gautier, M. Altorio, R. Geiger}%
\email{remi.geiger@obspm.fr}
\author{A. Landragin}
\affiliation{LNE-SYRTE, Observatoire de Paris-Universit\'e PSL, CNRS, Sorbonne Universit\'e,  61 avenue de l'Observatoire, 75014 Paris, France.}
\date{\today}%

\begin{abstract}
Multi-loop matter-wave interferometers are essential in quantum sensing to measure the derivatives of physical quantities in time or space. 
%They are realized by stacking several mirror stages, but the finite efficiency of the matter-wave mirrors creates spurious paths which scramble the signal of interest. 
Because multi-loop interferometers require multiple  reflections, imperfections of the matter-wave mirrors create spurious paths that scramble the signal of interest.
Here we demonstrate a method of adjustable momentum transfer that prevents the recombination of the spurious paths in a double-loop atom interferometer aimed at measuring rotation rates. We experimentally study the recombination condition of the spurious matter waves, which is quantitatively supported by a model accounting for the coherence properties of the atomic source. We finally demonstrate the effectiveness of the method in building a cold-atom gyroscope with a single-shot acceleration sensitivity suppressed by a factor of at least 50. Our study will impact the design of multi-loop atom interferometers that measure a single inertial quantity.
\end{abstract}

\maketitle

%%************************************

%%%********************
\label{intro}

Matter-wave interference is a central concept of quantum mechanics with a myriad of applications making use of electrons \cite{Missiroli1981}, neutrons~\cite{Rauch2000}, or atoms and molecules \cite{Cronin2009}. Examples of applications range from  bacteria characterization \cite{Dunin-Borkowski1998} and bio-molecular analysis \cite{Arndt2015}, to fundamental physics tests \cite{Abele2012} and accurate inertial sensing \cite{Geiger2020}.
%In most cases, a required high degree of control over the interference conditions and the precision of a measurement rely on the interference of two waves, with a sinusoidal fringe pattern providing direct access to the phase shift. 
In most cases, the  signal of interest can be detected as it shifts the phase of a sinusoidal interference fringe pattern of two partial waves.
However, the presence of auxiliary interferometic loops due to the imperfection of the mirrors  results in a multiple-wave interference, which reduces the interference contrast and the phase measurement accuracy.
%In most cases, the external effect of interest can be detected as its shifts the phase of a sinusoidal interference fringe pattern of two partial waves. However, if more than two branches contribute, they can reduce the interference contrast and the phase measurement accuracy.

Light-pulse atom interferometers employ a train of laser pulses that split, deflect and recombine the atomic waves to enclose a single loop Mach-Zehnder interferometer, in the simplest case. Here the light pulses act as atom optical beam splitters and mirrors, respectively.
Oftentimes, one may be interested in field derivatives rather than the fields themselves (e.g. gradients of the gravitational field or curvature of a magnetic field), or in a selective measurement in a given  frequency band. This is realized with interferometers consisting of several loops \cite{Clauser1988, McGuirk2002}, realized by multiple deflection of the matter-waves with additional mirrors - a technique analogous to the multi-pulse magnetic resonance spectroscopy \cite{Carr1954}.
%Accessing spatial derivatives of physical quantities (e.g. gradients of magnetic or gravitational fields) or a selective measurement of one among several contributions often requires interferometers consisting of several loops \cite{Clauser1988, McGuirk2002}, realized by multiple deflection of the matter-waves with additional mirrors - a technique analogous to the multi-pulse magnetic resonance spectroscopy \cite{Carr1954}. 

%The finite transmission of the atomic mirrors leads to the appearance of spurious leaked matter waves, which, in presence of additional mirrors, become re-directed and eventually form closed interferometric loops, degrading the two-wave nature of the interferometer \cite{Stockton2011}. Understanding and controlling the recombination of these spurious paths is intimately linked to the coherence of the matter-wave source, and requires a tailored design of the interferometric sequence and atomic mirrors.

The atom-optics  relies on coherent atom-light interaction, whose efficiency is limited by the homogeneity of the effective Rabi coupling that depends on the local laser intensity and velocity of the atom. The challenge arises when the non-zero transmission of the atomic mirrors  leads to leakage of the matter-waves, which are re-directed by subsequent mirrors and eventually form undesired additional interferometer loops, thus degrading the two-wave nature of the interferometer \cite{Stockton2011}. Understanding and controlling the recombination of these spurious paths is intimately linked to the coherence of the matter-wave source, and requires a tailored design of the interferometric sequence and atomic mirrors.

%In this Letter, we report on a method which prevents the recombination of the spurious paths in multi-loop cold-atom interferometers, both in position and in momentum space.  Our method uses atomic mirrors transferring an adjustable momentum to the atom. The high degree of control provided by this method, compared to other techniques in matter-wave interferometry, enables the detailed, though general, study of the process of the recombination of wave-packets. Finally, we show that the method of adjustable momentum transfer (AMT) allows for building a pure-rate gyroscope, as theoretically proposed in Ref.~\cite{Dubetsky2006}. Our results can be generalized to  atom-interferometric sensors of arbitrary multi-loop architectures. 

In this Letter, we report on a method which prevents the recombination of spurious paths in multi-loop cold-atom interferometers using mirrors that transfer an adjustable momentum to the atom. The high degree of control of this method, compared to other techniques in matter-wave interferometry, enables a detailed study of the recombination of wave-packets. We show that the method of adjustable momentum transfer (AMT) allows for building a pure-rate gyroscope (i.e. fully sensitive to rotation rate and insensitive to acceleration), as  proposed in Ref.~\cite{Dubetsky2006}. Our result can be generalized to atom-interferometer sensors of arbitrary multi-loop architectures.

We implement the AMT method in a double-loop atom interferometer aimed at measuring rotation rates and described in Refs.~\cite{Dutta2016,Savoie2018}. In short, we laser-cool Cesium atoms in a single internal state $|F=4\rangle$ to the temperature of 1.8~$\mu$K, and launch them vertically using moving molasses in an atomic fountain.
The atom optics employ stimulated Raman transitions at 852~nm that couple the $|F=3\rangle$ and $|F=4\rangle$ internal states with two counter-propagating laser fields of wave-vectors $\vec{k}_{3}$ and $\vec{k}_{4}$, imparting a momentum $\hbar\vec{k}_{\rm{eff}}=\hbar(\vec{k}_{3}-\vec{k}_{4})$ to the diffracted part of the wave-packet \cite{KasevichChu1991}. The interferometric sequence comprising four Raman laser pulses of $\pi/2, \pi, \pi, \pi/2$ Rabi angles, forms a symmetric double-loop interferometer. 
%The resulting accumulated phase difference is read out from the probability of transition between the two internal states using fluorescence detection and internal-state labeling of the output ports of the interferometer \cite{Borde1989}. 
Since the momentum  of the atoms is entangled with their internal state, the accumulated atomic phase difference is read out from the final population difference of the two hyperfine states, as determined by fluorescence detection \cite{Borde1989}.

\begin{figure}
	\centering
	\includegraphics[width=0.48\textwidth]{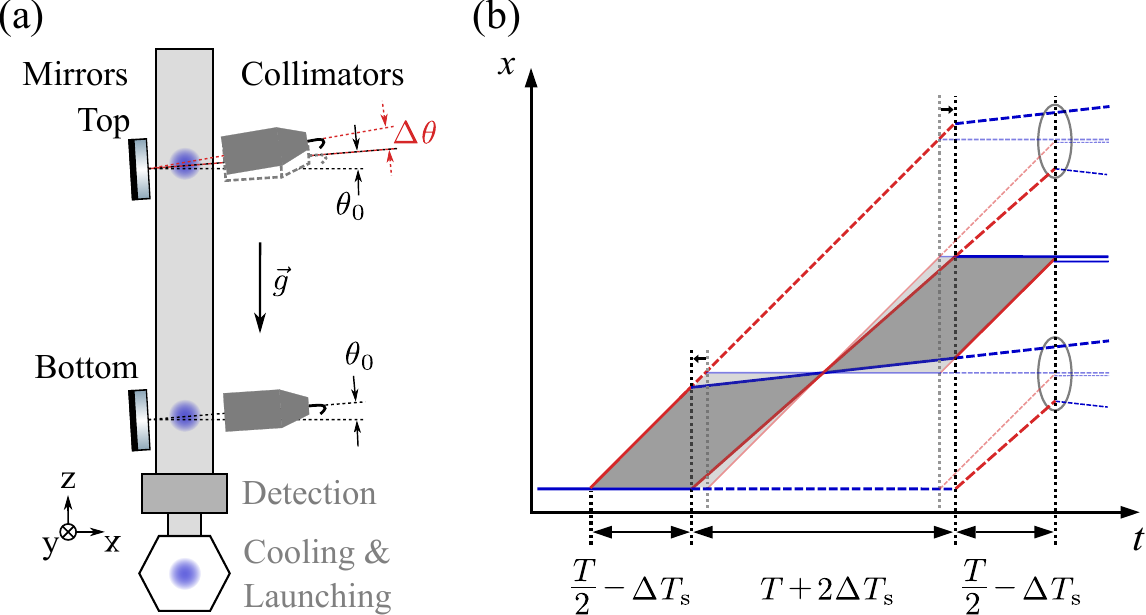}
	\caption{(a) Schematics of the cold-atom gyroscope sensor. The angular tilt $\Delta\theta$ of the top collimator allows for adjusting the effective momentum transfer of the $\pi$-pulses. 
	(b) Space-time diagram (not to scale) of the time-symmetric four-pulse interferometric sequence in the original equal-$\keff$ (thin half-transparent lines) and AMT (thick full lines) cases. Red (blue) color labels $F=3$ ($F=4$) internal state of the atoms. Solid (dashed) trajectories correspond to the main (spurious) interferometers. For equal-$\keff$ (AMT) sequence: the vertical dashed gray (black) lines indicate the timings of the pulses; light- (dark-) gray areas highlight the two loops of the main interferometer. Gray ovals mark the spatial separation of the spurious wave-packets at the last pulse in the AMT sequence. For clarity, we show only the output ports labeled by the $F=4$ state. 
		%Small horizontal arrows indicate the asymmetric ($\Delta T_{\rm{a}}$ in (c)) and symmetric ($\Delta T_{\rm{s}}$ in (d)) timing shifts of the middle pulses.
	}
	\label{fig:scheme_and_sequences}
\end{figure}

Two pairs of retro-reflected Raman beams interact with the atomic cloud at different height as shown in Fig.~\ref{fig:scheme_and_sequences}(a). 
The two mirrors are parallel to each other to better than $0.4~\mu$rad (see Ref.~\cite{Altorio2020} for the alignment procedure).
The normal to the mirrors, which sets the direction of the effective wave-vector $\vec{k}_{\rm{eff}}$, is inclined by an angle $\theta_{0}=3.8\degree$ with respect to the horizontal direction $\hat{x}$ (perpendicular to gravity), in order to lift the degeneracy between the $\pm\hbar\keff$ transitions owing to the Doppler effect.
%Two pairs of Raman beams access the interrogation region from two collimators as shown in Fig.~\ref{fig:scheme_and_sequences}(a). Each pair is retro-reflected by a mirror. The mirrors are parallel to each other to better than $0.2~\mu$rad \cite{Altorio2020} and tilted by an angle $\theta_{0}=3.8\degree$ with respect to the horizontal direction $\hat{x}$, in order to lift the degeneracy between the $\pm\hbar\keff$ transitions owing to the Doppler effect. 
The top collimator can be further inclined by a small adjustable angle $\Delta\theta\lesssim 20$~mrad, leading to a reduced modulus of the effective Raman wave-vector  of the top beam, $k_{\rm{eff}}^{\rm{(T)}}$, with respect to the bottom one, $k_{\rm{eff}}^{\rm{(B)}}\equiv k_{\rm{eff}}$, without changing its direction:
\begin{equation}\label{eq:deltak_epsilon_convention}
\begin{aligned}
k_{\rm{eff}}^{\rm{(B)}} - k_{\rm{eff}}^{\rm{(T)}} & = \epsilon k_{\rm{eff}}  \approx \frac{\Delta\theta^{2}}{2} k_{\rm{eff}}. 
\end{aligned} 
\end{equation}
This scheme, where the two wave-vectors are not equal (as theoretically studied in Ref.~\cite{Dubetsky2017amz} in the context of recoil frequency measurements) allows us to reach the necessary change in the momentum transfer to prevent the recombination of spurious paths in multi-loop interferometers with cold-atom sources. Adjusting the momentum transfer could also be realized by shifting the frequencies of the lasers, as proposed in Ref.~\cite{Roura2017} and implemented in Refs.~\cite{DAmico2017,Overstreet2018,Caldani2019} to reduce systematic errors in single-loop gravity sensors. 
%This approach would require an inconveniently large frequency change of tens of GHz, which makes it impractical in our case.
However, this would require frequency changes of tens of GHz, which makes it impractical, here.

In the traditional double-loop sequence \cite{Stockton2011,Dutta2016} (Fig.~\ref{fig:scheme_and_sequences}(b), thin half-transparent lines), the four Raman laser pulses are separated by time intervals $T/2$, $T$ and $T/2$, with $T=400$~ms . 
The time symmetry of this sequence with respect to the apogee of the atomic trajectory (crossing of the two loops at $t=t_{1}+T$, $t_{1}$ being the timing of the first pulse with respect to the launch) leads to a vanishing sensitivity to constant linear acceleration, which is required to build a pure-rate gyroscope.
However,  two spurious Ramsey-Bord\'{e}-like~\cite{Borde1984} interferometers (thin dashed lines in Fig.~\ref{fig:scheme_and_sequences}(b)) recombine simultaneously with the main one and, having different inertial sensitivity, impair the signal of interest.
Distinguishing the spurious interferometers from the main one would require a position-sensitive detector (along the $\hat{x}$ direction)  and an atomic source with sub-recoil temperature, which  would add complexity to the sensor architecture.

To circumvent this problem, one may apply a small asymmetric time shift of the mirror pulses (both pulses delayed or advanced by $\Delta T_{\rm{a}}$) \cite{Stockton2011}, inducing sufficient spatial separation of the spurious wave-packets while barely modifying the rotation-rate sensitivity of the main interferometer. Braking its time symmetry, nonetheless, imbalances the space-time areas of the two loops, which causes a sensitivity to  constant linear acceleration.
%barely modifies the rotation-rate sensitivity of the main linterferometer but
%suppressing the recombination of both spurious interferometers as soon as the spatial separation between the wave-packets exceeds the coherence length of the atoms. Braking the time symmetry, however, creates an imbalance between the space-time areas of the two loops of the main interferometer, which causes residual sensitivity to linear constant acceleration. 

The thick solid trajectory lines in Figure~\ref{fig:scheme_and_sequences}(b) show the interferometer sequence using AMT as explored in this Letter.  In order to close the main interferometer, the reduction of the momentum transfer at the $\pi$-pulses governed by Eq.~\eqref{eq:deltak_epsilon_convention} is compensated with their shift in time (see Fig. \ref{fig:scheme_and_sequences}(b)) of 
%changing the timings of the pulses (see Fig. \ref{fig:scheme_and_sequences}(b)) by   
%For a given value of $\epsilon$, we derive the expression for the symmetric time shift $\Delta T_{s}$ (applied now to both middle pulses) which allows for the perfect recombination at the fourth pulse as:
\begin{equation}
\Delta T_{s}=\frac{T\epsilon}{2(1-\epsilon)}
\label{eq:dT_dk_connection}
\end{equation} 
%where both $\Delta T_{\rm{s}}$ and $\epsilon$ are positive values.
%The new degree of freedom provided by AMT allows us to retain the original time symmetry of the main interferometer and prevents the recombination of the spurious wave-packets in both position and momentum. We will focus in the following on understanding the mechanism of recombination of the spurious interferometers and its link with the coherence of the atom source.  
The new degree of freedom provided by AMT allows us to retain the original time symmetry of the main interferometer and to prevent the recombination of the spurious wave-packets, on which we will focus in the following. 
\begin{figure*}
	\centering
	\includegraphics[width=0.98\linewidth]{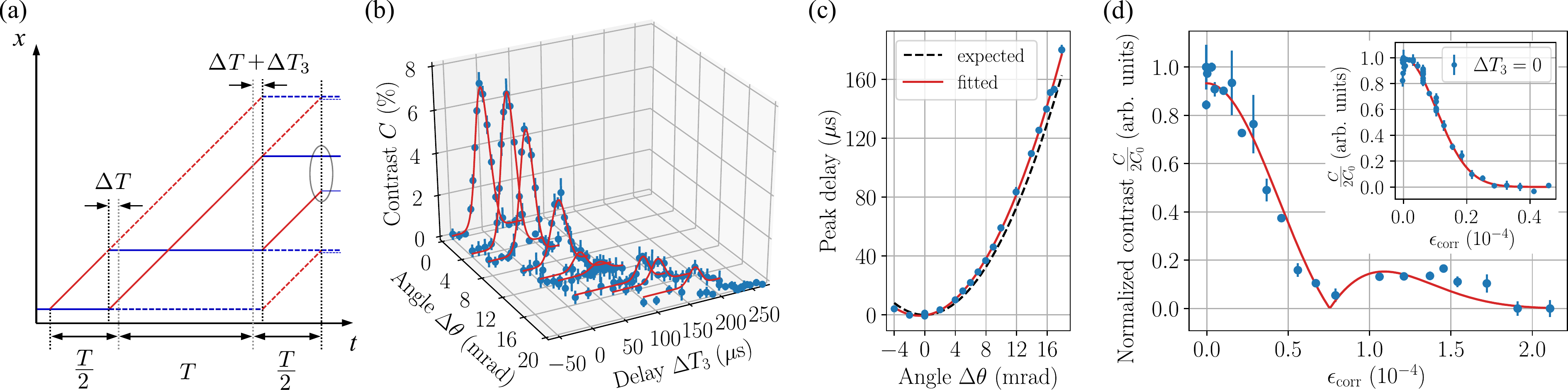}
	\caption{
		(a) Space-time diagram of the interferometer to introduce the definitions of the fixed $\Delta T=40~\mu$s and the variable $\Delta T_3$. (b) Peak-peak contrast of the spurious interferometers (blue dots) as a function of the third pulse delay $\Delta T_{3}$, for a set of angles $\Delta\theta$, and Gaussian fits (solid red lines) to the corresponding data. (c) Fitted values of $\Delta T_{3}$ yielding maximum contrast for the probed values of $\Delta\theta$. Dashed black line is the expectation of $\Delta T_{3}=2\epsilon(T+t_{1})$, solid red line is the fit to the data accounting for initial angular offsets $\Delta\theta_{0z}$, $\Delta\theta_{0y}$. (d) Normalized fitted peak contrast for the probed values of $\epsilon_{\rm{corr}}=\frac{1}{2}((\Delta\theta - \Delta\theta_{0z})^{2} - \Delta\theta_{0y}^{2})$. The solid red line is the fit with Eq.~\ref{eq:parasitic_contrast} with $\sigma_r$ and $C_{0}$ as free parameters (see text). Inset: contrast decay for $\Delta T_{3}=0$. The solid red line is the expectation for the measured value of $\sigma_{v}$ and the fitted value of $\sigma_{r}$.}
%		(a) Space-time diagram of the interferometer to introduce the definitions of the fixed $\Delta T=40~\mu$s and the variable $\Delta T_3$.
%		(b) Peak-peak contrast of the spurious interferometers (blue dots) as a function of the third laser pulse delay $\Delta T_{3}$, for different AMT angles $\Delta\theta$, and Gaussian fits to the corresponding data (solid red lines). 
%		%The black squares (dashed black line) are the data (Gaussian fit) of the main loop for $\Delta\theta=0$. 
%		(c) Top: fitted  values of $\Delta T_{3}$ yielding maximum contrast of the spurious interferometers for different values of $\Delta\theta$. Solid black line is the theoretical expectation of $\Delta T_{3}=2\epsilon(T+t_{1})$, dashed red line is the fit to the data accounting for initial angular offsets $\Delta\theta_{x0}$, $\Delta\theta_{y0}$. Bottom: residual with respect to the calculated (black squares) and fitted (red dots) curves. 
%		(d) Normalized fitted peak contrast for the probed values of $\epsilon_{\rm{calc}}=\frac{1}{2}((\Delta\theta - \Delta\theta_{x0})^{2} - \Delta\theta_{y0}^{2})$. The solid red line is the fit with Eq.~\ref{eq:parasitic_contrast} with $\sigma_r$ and $C_{0}$ as free parameters (see text). Inset: contrast decay for  $\Delta T_{3}=0$. The solid red line is the expectation for the  measured value of $\sigma_{v}$ and the fitted value of $\sigma_{r}$.}
	\label{fig:parasitic_interferometers}
\end{figure*}

In order to sufficiently separate the recombination time of the spurious interferometers from that of the main one (see Supplemental Material, section S3 \cite{SupMat}), we shift the mirror pulses by a fixed time interval $\Delta T=40~\mu$s as shown in Figure~\ref{fig:parasitic_interferometers}(a).
We also deliberately enhance the amplitudes in the spurious branches by changing the Rabi angles of these pulses from $\pi$ to $\pi/2$, thus making the mirrors half-transparent. 
Finally, we introduce a controlled variable delay of the third pulse, $\Delta T_{3}$ (see Fig.~\ref{fig:parasitic_interferometers}(a)), that allows us to probe the efficiency of the recombination of the spurious interferometers. 
%To specifically study the recombination of the spurious loops independently from the main two-loop interferometer, we shift in time the second and third pulses by a fixed value $\Delta T=40~\mu$s (Fig.~\ref{fig:parasitic_interferometers}(a)). This value ensures sufficient separation in recombination time between the spurious and the main interferometers ,such that the wings of the latter do not affect the signal of the spurious loops (see Supplemental Material, section S3 \cite{SupMat}).
%To enhance that signal, we also deliberately increase the amplitudes in the spurious branches by changing the duration of the second and third pulses from $\pi$ to $\pi/2$, thus making the mirrors half-transparent. 
%We introduce a variable time delay of the third laser pulse, $\Delta T_{3}$ (see Fig.~\ref{fig:parasitic_interferometers}(a)), which allows us to probe the efficiency of the recombination of the spurious interferometers.

In Figure~\ref{fig:parasitic_interferometers}(b) we present the evolution of the contrast of the spurious interferometers as we gradually transform the sequence with increasing value of the angle $\Delta\theta$. For each angle, we 
%follow the spurious interferometer contrast 
probe the spurious signal by scanning the value of $\Delta T_{3}$ (blue dots). At $\Delta\theta=0$, we find the maximum of the contrast, as expected, around $\Delta T_{3} = 0~\mu\mathrm{s}$. 
%For completeness, we verify that the main interferometer (black squares) is centered around $\Delta T_{3} = 0~\mu$s, sufficiently distant such that its wings do not affect the spurious signal.
We observe a reduction of the maximum contrast 
%of the spurious interferometers 
while increasing $\Delta\theta$ towards an almost full suppression around $\Delta\theta=12$~mrad, followed by a clear revival and a final decay at large angles. To connect the observed contrast behavior with the coherence properties of the cold-atom source, we derive the phase shifts of the bottom (B) and top (T) spurious interferometers (see Supplemental Material, section S1 \cite{SupMat}) as: 
%for a test wave-packet having initial (at launch) classical position $\vec{r}_{0}=\vec{r}(t=0)$ and velocity $\vec{v}_{0}=\vec{v}(t=0)$:  
%In Figure~\ref{fig:parasitic_interferometers}(b) we present the evolution of the contrast of the spurious interferometers as we gradually transform the sequence with increasing value of the angle $\Delta\theta$. For each angle, we follow the spurious interferometer contrast by scanning the value of $\Delta T_{3}$ (blue dots). At $\Delta\theta=0$, we find the maximum of contrast, as expected, around $\Delta T_{3} = 0~\mu\mathrm{s}$. 
%For completeness, we verify that the main interferometer (black squares) is centered around $\Delta T_{3} = 0~\mu$s, sufficiently distant such that its wings do not affect the spurious signal.
%We observe a reduction of the maximum contrast of the spurious interferometers while increasing $\Delta\theta$ towards an almost full suppression around $\Delta\theta=12$~mrad, followed by a clear revival and a final decay at large angles. To explain these observations, we connect the contrast of the spurious interferometers with the coherence properties of the cold-atom source, by deriving the interferometer phase shifts. The calculation details are given in \cite{SupMat} (section S1) and the result for the bottom (B) and top (T) spurious interferometers  is  
\begin{equation}
\label{eq:parasitic_phase_shift}
\begin{aligned}
\Delta\Phi^{\rm{(B)}} & = \Delta\Phi_r(r_0) + \Delta\Phi_v(v_0) + \Delta\Phi^{\prime}-\omega_{R}T\epsilon\\
\Delta\Phi^{\rm{(T)}} & = \Delta\Phi_r(r_0) + \Delta\Phi_v(v_0) + \Delta\Phi^{\prime}+\omega_{R}T\epsilon,\\
\end{aligned} 
\end{equation}
where $\omega_{R}\equiv\frac{\hbar\keff^2}{2m}$ is the two-photon recoil frequency and
\begin{equation}
\label{eq:dphi_r_v}
\begin{aligned}
\Delta\Phi_r({r_{0}}) & = 2\vec{k}_{\rm{eff}}\cdot \vec{r}_{0}\epsilon\\
\Delta\Phi_v({v_{0}}) & = \vec{k}_{\rm{eff}}\cdot\vec{v}_{0}\big(2\epsilon\left(T+t_{1}\right)-\Delta T_{3}\big).
\end{aligned} 
\end{equation}
We express the phase shifts as a sum of four distinct terms. The first two terms, $\Delta\Phi_r(r_{0})$ and $\Delta\Phi_v(v_{0})$, depend on the initial (at launch) position $\vec{r}_{0}=\vec{r}(t=0)$ and velocity $\vec{v}_{0}=\vec{v}(t=0)$ of a given atom in the reference frame of the center of mass of the atomic cloud.
%, and are thus governed by the statistical properties of the atomic source. 
The third term, $\Delta\Phi^{\prime}$, incorporates the  inertial contribution due to acceleration and the common recoil phase shift. 
%of the Ramsey-Bord\'e interferometers. 
The last term constitutes a relative dephasing of the two spurious interferometers, which increases with $\epsilon$. 

The contrast of the spurious interferometric signal, $C(\epsilon, \Delta T_{3})$, is given by the (incoherent) sum of the intensities from both interferometers, averaged over the initial statistical velocity and position distributions of the atomic source. We assume uncorrelated Gaussian velocity and position distributions, respectively characterized by the standard deviations $\sigma_{v}$ and $\sigma_{r}$, and obtain (see Supplemental Material, section S2 \cite{SupMat}):  
\begin{equation}\label{eq:parasitic_contrast}
%\begin{aligned}
%\frac{C(\epsilon, \Delta T_{3})}{2C_{0}}=\big|\cos\left(\omega_{R}T\epsilon\right)\big| \exp{\left\{-\frac{\Delta\Phi_{r}^{2}+\Delta\Phi_{v}^{2}}{2}\right\}}\bigg|_{\substack{r_{0}=\sigma_{r}\\v_{0}=\sigma_{v}}}
%\frac{C(\epsilon, \Delta T_{3})}{2C_{0}}&=\big|\cos\left(\omega_{R}T\epsilon\right) \big|\times\\
%&\times \exp{\left\{-\frac{\Delta\Phi_{r}(\sigma_r)^{2}+\Delta\Phi_{v}(\sigma_v)^{2}}{2}\right\}}
\frac{C(\epsilon, \Delta T_{3})}{2C_{0}}=\big|\cos\left(\omega_{R}T\epsilon\right) \big|e^{-\frac{1}{2}(\Delta\Phi_{r}(\sigma_r)^{2}+\Delta\Phi_{v}(\sigma_v)^{2})},
%C(\epsilon, \Delta T_{3})=2C_{0}\big|\cos\left(\omega_{R}T\epsilon\right) \big|\times e^{-\frac{1}{2}(\Delta\Phi_{r}(\sigma_r)^{2}+\Delta\Phi_{v}(\sigma_v)^{2})}
%\end{aligned}
\end{equation}
where $C_{0}$ is the maximum mean contrast. The oscillating term reflects the recoil-originated dephasing between the two spurious interferometers, while the exponential suppression factor 
%contains contributions associated with 
highlights the role of the finite spatial and momentum spread in the cold-atom source.
%spatial ($r_0$) and temporal ($v_0$) coherence of the cold-atom source.  

%The finite spatial coherence reduces the contrast by a  factor that is solely given by the initial cloud size, $\sigma_{r}$. 
The effect of the finite velocity spread on the contrast can be fully eliminated by the proper choice of $\Delta T_{3}=2\epsilon(T+t_{1})$ (from Eq.~\eqref{eq:dphi_r_v}), which defines the recombination in momentum space and therefore yields  the maximum of contrast. 
This expectation for $\Delta T_3$, shown by the dashed black line in Figure~\ref{fig:parasitic_interferometers}(c), qualitatively  matches the data (blue dots). 
A quantitative agreement is obtained by accounting for the initial angular mismatch between the collimators in both $z$ (vertical) and $y$ (horizontal) directions via $\epsilon_{\rm{corr}} = ((\Delta\theta-\Delta\theta_{0z})^{2}-\Delta\theta_{0y}^{2})/2$~\cite{NoteTheta0y}. Fitting the data with $\Delta T_{3}=2\epsilon_{\rm{corr}}(T+t_{1})$ (solid red line in Fig.~\ref{fig:parasitic_interferometers}(b)) reveals small angular offsets $\Delta\theta_{0z}=-0.68(5)$~mrad and $\Delta\theta_{0y}=1.04(19)$~mrad, which are compatible with the inaccuracy of the initial manual tuning of the collimator of about 1~mrad.

In Figure~\ref{fig:parasitic_interferometers}(d) we plot the values of fitted maximum contrast of the spurious interferometers normalized to the maximum value among all the datasets, for different values of $\epsilon_{\rm{corr}}$. The overall trend, including the  zero and the revival, is well reproduced by the fit (solid red line) with the model of Eq.~\eqref{eq:parasitic_contrast} accounting for $\Delta\Phi_v=0$ (recombination in momentum space), with  $C_{0}$ and $\sigma_{r}$ as free parameters. 
The fitted value of $\sigma_{r}=0.51(2)$~mm sets the realistic scale for the spatial extent of the atomic cloud.
A non-Gaussian actual cloud shape  might be the cause of the slight mismatch around $\epsilon_{\rm{corr}}\simeq 1.6\times 10^{-4}$. 

The inset of Figure~\ref{fig:parasitic_interferometers}(d) shows the contrast decay in the case of $\Delta T_{3}=0$ (where $\Delta\Phi_v\neq 0$), which is driven by the finite velocity spread of the source and happens on a much faster $\epsilon$-scale. 
This behavior does not depend on the specific value of $\Delta T=40 \ \mu$s, and is thus also applicable to the case of $\Delta T=\Delta T_{\rm{s}}$ (Fig.~\ref{fig:scheme_and_sequences}(b)). The complete suppression of the signal of the spurious loops in the time-symmetric AMT sequence therefore  happens on a scale of $\epsilon \simeq 4\times 10^{-5}$. 
This data is well matched by the expected behavior of Eq.~\eqref{eq:parasitic_contrast} (solid red line), with the value of $\sigma_{v}$ extracted from the widths of the peaks in the panel (b) (see Supplemental Material, section S3 \cite{SupMat}). 
%This value of $\sigma_v$ is consistent with the velocity selectivity of the Raman transitions.

%The solid red line represents the expected behavior of Eq.~\eqref{eq:parasitic_contrast} with the measured value $\sigma_{v}=6.4(8)$~mm/s (extracted from the time-widths of the peaks in (b), see Supplemental Material, section S3 \cite{SupMat}) and matches well the data. 
%This value of $\sigma_v$ is consistent with the velocity selectivity of the Raman transitions.

\begin{figure}[t]
	\centering
	\includegraphics[width=0.48\textwidth]{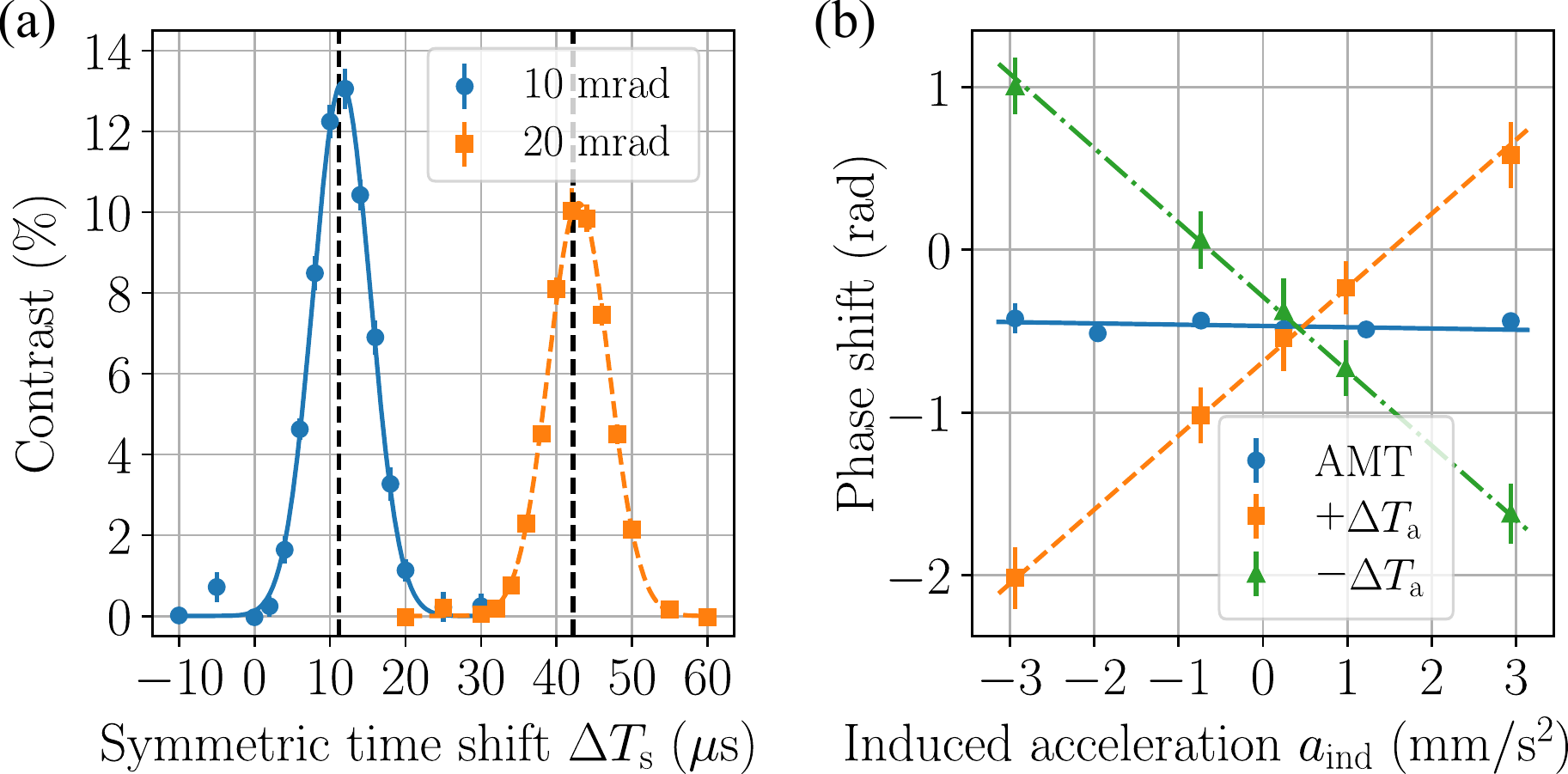}
	\caption{(a) Contrast of the main interferometer in the AMT configuration as a function of symmetric time shift $\Delta T_{\rm{s}}$, for two values of the 
%	collimator tilt 
	angle $\Delta\theta$. The solid blue and dashed orange lines are empiric Gaussian fits to the data. The vertical dashed lines mark the expected center positions. (b) Phase shift as a function of induced acceleration (see text), in the AMT case with $\Delta\theta=20$~mrad and $\Delta T_{\rm{s}}=42.9~\mu$s (blue dots), and in the asymmetric case with $\Delta T_{\rm{a}}=\pm40~\mu$s (orange squares and green triangles). Solid blue, dashed orange and dash-dotted green lines are linear fits to the data.} 
	\label{fig:main_recombination_tilt}
\end{figure}

We now focus on the main (double-loop) interferometer in the time-symmetric AMT configuration of Figure~\ref{fig:scheme_and_sequences}(b). 
The promised insensitivity to the dc-acceleration of this sequence, in practice, relies on the ability to accurately meet the condition of Eq.~\eqref{eq:dT_dk_connection}. 
%This, in turn, is linked with the precise knowledge of the laser pulse timings and taking into account the finite duration of the pulses. 
In Figure~\ref{fig:main_recombination_tilt}(a), we probe the recombination of the main interferometer for the applied values of $\Delta\theta=10$~mrad and 20~mrad by scanning the time shift $\Delta T_{\rm{s}}$ and recording the peak-peak contrast. We find the peak centers at $11.6(1)~\mu\rm{s}$ and $42.9(1)~\mu\rm{s}$, very close to their respective expectation of Eq.~\eqref{eq:dT_dk_connection} with $\epsilon=\epsilon_{\rm{corr}}$ at $11.2~\mu\rm{s}$ and $42.2~\mu\rm{s}$.
%$f(x)=A\exp((x-c)^2/2\sigma^2)$
%We fit both peaks with empiric Gaussian profiles and extract the peak centers at $11.56(7)~\mathrm{\mu s}$ and $42.92(4)~\mathrm{\mu s}$, very close to their respective expectation of Eq.~\eqref{eq:dT_dk_connection} with $\epsilon=\epsilon_{\rm{corr}}$ at $11.29~\mathrm{\mu s}$ and $42.38~\mathrm{\mu s}$.

We choose the AMT arrangement with $\Delta\theta=20$~mrad and $\Delta T_{\rm{s}}=42.9~\mathrm{\mu s}$ to  verify the insensitivity of the  main interferometer to the linear dc-acceleration. We induce an additional acceleration along the $\keff$-direction via controlled tilt of the sensor in x-z plane by a small angle $\beta$ such that $a_{\mathrm{ind}}=g\cos{\theta_0}\sin{\beta}$, and measure the corresponding phase shift (blue dots in Fig.~\ref{fig:main_recombination_tilt}(b)). 
%The result is shown by the blue dots in Fig.~\ref{fig:main_recombination_tilt}(b). 
The fitted residual linear slope $d\Phi/da_{\mathrm{ind}}=0.4(8.5)~\mathrm{rad/(m}\cdot\mathrm{s}^{-2})$ (blue line in Fig.~\ref{fig:main_recombination_tilt}(b)) is compatible with zero within the error bar. 
For comparison, we perform 
%a measurement of the effect of induced dc-acceleration 
an identical measurement in the asymmetric configuration \cite{Stockton2011,Dutta2016} where both mirror pulses are advanced ($\Delta T_{\mathrm{a}}=40~\mu\mathrm{s}$, orange squares) or delayed ($\Delta T_{\mathrm{a}}=-40~\mu\mathrm{s}$, green triangles). For this configuration, we extract the respective dc-acceleration-sensitivity slopes of  $446(8)~\mathrm{rad/(m}\cdot\mathrm{s}^{-2})$ and $-448(5)~\mathrm{rad/(m}\cdot\mathrm{s}^{-2})$ matching within 5\% the expectation of $d\Phi/da_{\mathrm{ind}}=2T\Delta T_{\mathrm{a}}\keff$. 
The ratio of the modulus of the slopes reflects a suppression of the acceleration-induced phase shift in a single measurement using the symmetric AMT configuration, as compared to the asymmetric one, by at least the factor of 50. 
%In practice, mitigation of slow variations of the acceleration requires successive measurements with alternating sign of $\Delta T_{a}$. 
%The gyroscope remains, however, sensitive to fluctuations of the linear acceleration~\cite{Savoie2018} at the frequency of the order of inverse of the measurement repetition rate.
% In practice, the static contribution from the acceleration can be suppressed by a ramp of the difference of frequencies between the two Raman lasers with a rate given by $\alpha=\vec{k}_{\rm{eff}}\vec{g}$ which compensates for the Doppler shift at the moment of light pulses' (a technique widely employed in atomic gravimetry). Futhermore, slow variations of the acceleration can be mitigated by successive measurements with alternating sign of $\Delta T_{a}$. The gyroscope remains, however, sensitive to fluctuations of the linear acceleration at the frequency of the order of inverse of the measurement repetition rate.The absolute value of the slopes for usual asymmetric configuration agrees with Eq.~\ref{eq:dPhi_acc} within 5.4\%, which we attribute to a slight inaccuracy of the tiltmeter (see Appendix~\ref{sec:AppendixD}). The remarkable crossing of the three fitted lines in a single point shows that the acceleration phase is well compensated for the usual asymmetric configuration with a proper choice of the frequency ramp value $\alpha$.

We finally consider the impact of the AMT technique on the phase shift of the gyroscope sensor, which is given by (see Supplemental Material, section S4 \cite{SupMat}):
%We finally consider the impact of the AMT technique on the phase shift of the gyroscope sensor. The  phase shift of the two-loop interferometer reads:
\begin{equation}\label{eq:main_phase_shift}
\begin{aligned}
\Delta\Phi = \frac{1}{2}\vec{k}_{\rm{eff}}(\vec{g}\times\vec{\Omega})T^{3}\left(1-\frac{2\epsilon}{3}\right)+\Delta\omega_{0}\frac{2T\epsilon}{(1-\epsilon)}.\\
\end{aligned} 
\end{equation}
The first term accounts for a correction to the gyroscope scale factor, as can be derived from the reduction of the physical (Sagnac) area of the interferometer, with $\vec{\Omega}$ being the rotation rate of the Earth. The second term (called hereafter clock shift) represents the sensitivity to the detuning ($\Delta\omega_0$) of the relative Raman laser frequencies from the resonance condition of the Raman transition at the apogee point. 

We measured the clock shift and confirmed the expected behavior of Eq.~\eqref{eq:main_phase_shift} (see Supplemental Material, section S4 \cite{SupMat}). By alternating measurements with $\pm \keff$, we could demonstrate a rejection of this clock shift by at least two orders of magnitude, yielding a residual sensitivity for $\Delta\theta=20$~mrad compatible with zero and below 10 mrad/kHz. 
The study of the  correction to the gyroscope scale factor (first term of Eq.~\eqref{eq:main_phase_shift}) goes beyond the scope of this Letter.

To conclude, we have demonstrated the method of adjustable momentum transfer in multi-loop atom interferometers,
%that efficiently suppresses spurious interferometers
that provides a controlled suppression of the spurious interferometric signals originating from the finite efficiency of atomic mirrors.
The observed variation of the contrast of the spurious interferometers, quantitatively supported by our model, revealed a fractional imbalance in momentum transfer of $\epsilon\simeq 4\times 10^{-5}$ to completely suppress the spurious signals, as given by the finite coherence of our atom source.
%The  study of the variation in contrast of the spurious interferometers, quantitatively supported by our model, revealed that an imbalance in momentum transfer of $\epsilon\simeq 4\times 10^{-5}$  completely suppresses the signal of the spurious loops, accounting for the coherence of our atom source. 
%Moreover, this study allowed us to highlight 
In addition, we discovered a remarkable configuration ($\omega_R T\epsilon=\pi/2$), where the spurious interferometers are in anti-phase due to their different recoil sensitivity. 
The AMT method allowed us to demonstrate a double-loop gyroscope with a highly suppressed sensitivity to constant linear acceleration. This holds particular interest for applications where the fluctuations of the rotation rate of the ground need to be discriminated from the linear translations, as for example, in the field of rotational seismology \cite{Hadziioannou2012}.

Our results pave the way for the design of sensors with  atomic sources of increased coherence or with more than two interferometric loops, where the problems associated with spurious paths are enhanced.
More generally, our work shows the possibility of tuning the sensitivity of multi-loop atom interferometers to a unique, chosen, physical quantity, which enables to extend the scope of atom interferometry to new domains.
This is crucial for multi-loop atom interferometers used as gravity gradiometers \cite{McGuirk2002,Perrin2019} and gyroscopes \cite{Canuel2006,Stockton2011,Dutta2016}, or proposed for  gravitational wave detection \cite{Hogan2011,Graham2016a,Canuel2018,schubert2019} or for measuring space-time curvature \cite{Marzlin1996}.

\begin{acknowledgments}
We thank Peter Wolf and Albert Roura for stimulating discussions and Franck Pereira dos Santos for careful reading of the manuscript.
We acknowledge the financial support from Ville de Paris (project HSENS-MWGRAV), FIRST-TF (ANR-10-LABX-48-01), Centre National d'Etudes Saptiales (CNES), Sorbonne Universit\'es (project SU-16-R-EMR-30, LORINVACC) and Agence Nationale pour la Recherche (project PIMAI, ANR-18-CE47-0002-01). L.A.S. was funded by Conseil Scientifique de l'Observatoire de Paris (PSL fellowship in astrophysics at Paris Observatory), M.A. and R. Gautier by the EDPIF doctoral school. 
\end{acknowledgments}

%\bibliography{NEK_paper_bib}

\begin{thebibliography}{33}
\bibitem [{Not()}]{NoteTheta0y}%
  \BibitemOpen
  \href@noop {} {}\bibinfo {note} {The minus sign in front of the
  $\Delta\theta_{0y}^{2}$ indicates that the bottom collimator is more tilted
  than the top one in the $y$ (horizontal) direction.}\BibitemShut {Stop}%
\end{thebibliography}
%\label{biblio}


\onecolumngrid
%\appendix
\newpage

\section*{SUPPLEMENTAL MATERIAL for \\  Tailoring multi-loop atom interferometers with adjustable momentum transfer}

\renewcommand{\thefigure}{S\arabic{figure}}
\renewcommand{\theequation}{S\arabic{equation}}
\setcounter{figure}{0}
\setcounter{equation}{0}

\section*{S1. Phase shifts of the spurious loops}
\label{sec:AppendixA}
\begin{figure}[ht]
	\centering
	\includegraphics[width=0.35\textwidth]{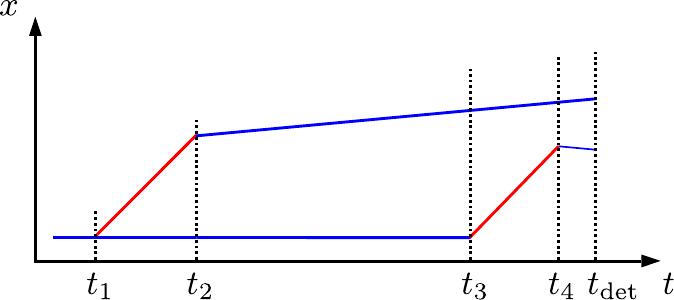}
	\caption{Sketch of the bottom spurious interferometer sequence until the detection moment. Vertical dashed lines indicate the four timings of the applied light pulses $\{t_{\rm{i}}\}$ and of the detection pulse. Red (blue) color labels $F=3$ ($F=4$) internal state of the atoms. For clarity, we show only the output port corresponding to the $F=4$ state.}
	\label{fig:A_scheme_bottom_echo_loop}
\end{figure}

Given a macroscopic separation of the two spurious interferometers of about 2.5~mm at the detection moment, they do not interfere between each other and may be considered separately. Below we derive the phase shift of the bottom spurious interferometer (Fig.~\ref{fig:A_scheme_bottom_echo_loop}) at the moment of detection. As the calculation for the top spurious loop is conceptually similar, we only provide the final result. 
In the following, we neglect the finite time length of the laser pulses (all pulses have an area of $\pi/2$ and time length of $10~\mu\mathrm{s}<<400~\mathrm{ms}=T$) and consider them applied at the time moments $t_{1} .. t_{4}$ (see Fig. A1), while $t=0$ moment corresponds to the launch of the atomic cloud:
\begin{equation}\label{eq:A1_timing_definition}
\begin{aligned}
t_{1} & = 114~\mathrm{ms}\\ 
t_{2} & = t_{1} + T/2 - \Delta T\\
t_{3} & = t_{1} + 3T/2 + \Delta T + \Delta T_{3}\\
t_{4} & = t_{1} + 2T\\
t_{\mathrm{det}} & = t_{1} + 2T + \Delta t_{\mathrm{det}},\\
\end{aligned}
\end{equation}     
where $\Delta T=40~\mu\mathrm{s}$ is the initial time shift that separates in time the recombination moments of the main and spurious inerferometers (see main text and additional data section below), $\Delta T_{3}$ is the delay of the third pulse and $\Delta t_{\mathrm{det}}=70$~ms is the time interval past the last laser pulse until the detection moment ($t_{\mathrm{det}}$).

The total interferometric phase  shift may be represented as a sum of three parts \cite{Wolf2011daa}:
%\cite{Wolf1999gua} 
\begin{equation}\label{eq:A1_full_parasitic_phase_shift}
\Delta\Phi = \Delta\Phi_{\mathrm{las}} + \Delta\Phi_{\mathrm{prop}} + \Delta\Phi_{\mathrm{sep}}\\ 
\end{equation}
Here $\Delta\Phi_{\mathrm{las}}$ is the laser phase which is imprinted onto the atomic wave-packet via interaction with Raman laser pulses; $\Delta\Phi_{\mathrm{prop}}$ is the free propagation phase difference accumulated along the paths; $\Delta\Phi_{\mathrm{sep}}$ is the phase shift arising from spatial separation of the two wave-packets at the moment of detection (interference), so-called separation phase. We take the convention for $\Delta\Phi$ being the phase shift of the upper branch minus the phase shift of the lower branch, and mark the related variables with u(l) subscripts.\\ 
\\
\noindent\textbf{\textit{Laser phase}} The laser phase reads as:
\begin{equation}\label{eq:A1_parasitic_laser_phase_initial}
\begin{aligned}
\Delta\Phi_{\rm{las}} & = (\varphi_{1}-\varphi_{2})_{\rm{u}}-(\varphi_{3}-\varphi_{4})_{\rm{l}}\\ 
\varphi_{1} & = \vec{k}_{\rm{eff}}\vec{r}_{\mathrm{u}}(t_{1}) - \int_{0}^{t_{1}}\omega_{\mathrm{eff}}(t)dt\\
\varphi_{2} & = (1-\epsilon)\vec{k}_{\rm{eff}}\vec{r}_{\mathrm{u}}(t_{2}) -  \int_{0}^{t_{2}}\omega_{\mathrm{eff}}(t)dt\\
\varphi_{3} & = (1-\epsilon)\vec{k}_{\rm{eff}}\vec{r}_{\mathrm{l}}(t_{3}) - \int_{0}^{t_{3}}\omega_{\mathrm{eff}}(t)dt\\
\varphi_{4} & = \vec{k}_{\rm{eff}}\vec{r}_{\mathrm{l}}(t_{4}) -  \int_{0}^{t_{4}}\omega_{\mathrm{eff}}(t)dt\\
\end{aligned}
\end{equation}
For completeness, we account here for the phase change due to the ramp of two-photon laser frequency to fulfill the resonance condition due to the Doppler shift: $\omega_{\mathrm{eff}}(t) = \omega_{0} - \alpha(t-T-t_1)$, where $\omega_{0} = \omega_{hf}+\hbar k_{\mathrm{eff}}^2 /2m$ expresses the resonance condition for the atom at rest, namely at the apogee point of trajectory at $t=t_{1}+T$. The ramp rate $\alpha$ is given by the projection of the gravity acceleration on the Raman beams in $\Delta\theta=0$ configuration: $\alpha=k_{\mathrm{eff}}g\sin(\theta_{0})$. We also note that we orient the X-Z plane of our sensor (plane in which the area of the loops opens) towards the geographic West thus zeroing any possible contribution from the Earth rotation rate $\vec{\Omega}$.   
 
We now consider an atom (wave-packet) with initial ($t=0$) classical velocity $\vec{v}_{0}$ and position $\vec{r}_{0}$ and express the position of the wave-packet at the relevant time moments, along the upper and lower branch: 
\begin{equation}\label{eq:A1_parasitic_coordinates}
\begin{aligned}
\vec{r}_{\mathrm{u}}(t_{1}) & = \vec{r}_{0} + \vec{v}_{0}t_{1} + \frac{1}{2}\vec{g}t_{1}^{2}\\
\vec{r}_{\mathrm{u}}(t_{2}) & = \vec{r}_{0} + \vec{v}_{0}t_{2} + \frac{1}{2}\vec{g}t_{2}^{2} + \frac{\hbar\vec{k}_{\mathrm{eff}}}{m}(t_{2}-t_{1})\\
\vec{r}_{\mathrm{l}}(t_{3}) & = \vec{r}_{0} + \vec{v}_{0}t_{3} + \frac{1}{2}\vec{g}t_{3}^{2}\\
\vec{r}_{\mathrm{l}}(t_{4}) & = \vec{r}_{0} + \vec{v}_{0}t_{4} + \frac{1}{2}\vec{g}t_{4}^{2} + \frac{\hbar\vec{k}_{\mathrm{eff}}}{m}(1-\epsilon)(t_{4}-t_{3})\\
\end{aligned}
\end{equation}
We plug these formulae into the equations~\ref{eq:A1_parasitic_laser_phase_initial} and, with the use of explicit timings (Eqs.~\ref{eq:A1_timing_definition}) and obtain the final result for the laser phase shift, where we neglect the terms of the orders of $O(\epsilon^{2}, \epsilon\Delta T/T, \epsilon\Delta T_{3}/T)$:  
\begin{equation}\label{eq:A1_parasitic_laser_phase_final}
\begin{aligned}
& \Delta\Phi_{\rm{las}} = \Delta\Phi_{\rm{r}_{0}} + \Delta\Phi_{\rm{v}_{0}} + \Delta\Phi_{\rm{g}+\alpha} + \Delta\Phi_{\rm{rec}}\\
& \Delta\Phi_{\rm{r}}(r_{0}) = 2\vec{k}_{\mathrm{eff}}\vec{r}_{0}\epsilon\\
& \Delta\Phi_{\rm{v}}(v_{0}) = \vec{k}_{\mathrm{eff}}\vec{v}_{0}(2\epsilon(T+t_{1})-\Delta T_{3})\\  
& \Delta\Phi_{\rm{g}+\alpha} = (\vec{k}_{\mathrm{eff}}\vec{g}-\alpha)\Big(\frac{3}{4}T^{2}-T\Delta T-\frac{3}{2}T\Delta T_{3}\Big)+ \vec{k}_{\mathrm{eff}}\vec{g}\Big(\epsilon\big(\frac{5}{4}T^{2} + 2Tt_{1}+t_{1}^2\big)-t_{1}\Delta T_{3}\Big)\\
& \Delta\Phi_{\rm{rec}}= -\frac{\hbar\keff^2}{m}\Delta T_{3}\\ 
\end{aligned} 
\end{equation}   
The terms $\Delta\Phi_{\rm{r}}(r_{0})$ and $\Delta\Phi_{\rm{v}}(v_{0})$ depend on the initial position and velocity of the wave-packet. 
%and should be averaged over the statistical distribution of an atomic ensemble to account for our detection scheme which does not discriminate position or velocity..  
%The recoil part $\Delta\Phi_{\rm{rec}}$ adds a small constant offset and, as it is invariant under the change of $\vec{k}_{\mathrm{eff}}$, cancels in the half-difference of $\pm\vec{k}_{\mathrm{eff}}$ signals. 
The last term $\Delta\Phi_{\rm{g}+\alpha}$ contains the dc-acceleration shift which is in the leading order compensated by the frequency ramp $\alpha$. 
%The interferometer, however, remains fully sensitive to the ac-acceleration which in practice produces many radians of phase shift, resulting in a random shot-to-shot sampling of the fringe.\\    
\\
\noindent\textbf{\textit{Free propagation phase}} The free propagation phase is given by the integrals of the Lagrangian along the two corresponding classical paths \cite{Storey1994tfp}:
\begin{equation}\label{eq:A1_parasitic_free_phase_initial}
\begin{aligned}
& \Delta\Phi_{\rm{prop}} = \int\limits_{u} L(t)dt - \int\limits_{\mathrm{l}} L(t)dt\\
& L(t) = \frac{1}{2}mv(t)^{2}-m\vec{g}\vec{r}(t)\\
\end{aligned} 
\end{equation} 
Considering as before an atom (wave-packet) with initial classical velocity $\vec{v}_{0}$ and position $\vec{r}_{0}$, we perform straightforward integration until the moment of detection and obtain (neglecting same higher-order terms as in the calculation for the laser phase): 
\begin{equation}\label{eq:A1_parasitic_free_phase_final}
\begin{aligned}
\Delta\Phi_{\rm{prop}} & = \vec{k}_{\mathrm{eff}}\vec{v}_{0}(\Delta T_{3}+2\epsilon(T+\Delta t_{\mathrm{det}})) + \frac{\hbar\keff^2}{2m}(\Delta T_{3}+\epsilon T)\\
\end{aligned} 
\end{equation}
\\
\noindent\textbf{\textit{Separation phase}} This contribution arises from the fact that one detects the interference between two wave-packets at a given location $\vec{r}$ in the detection region which has certain distances from the positions of the two classical trajectory points $\vec{r}_{\mathrm{u}}(t_{\mathrm{det}})$ and $\vec{r}_{\mathrm{l}}(t_{\mathrm{det}})$. The general expression for this phase shift reads \cite{Johnson2011lba}:
\begin{equation}\label{eq:A1_parasitic_sep_phase_initial}
\begin{aligned}
\Delta\Phi_{\rm{sep}} & = \frac{1}{\hbar}(\vec{p}_{\mathrm{u}}(\vec{r}-\vec{r}_{\mathrm{u}})-\vec{p}_{\mathrm{l}}(\vec{r}-\vec{r}_{\mathrm{l}})) = \frac{1}{\hbar}(-\vec{p}_{c}\vec{\Delta r}+\vec{\Delta p}(\vec{r}-\vec{r}_{c}))\\
\vec{p}_{c} & = \frac{\vec{p}_{\mathrm{u}} + \vec{p}_{\mathrm{l}}}{2}, \vec{\Delta p} = \vec{p}_{\mathrm{u}} - \vec{p}_{\mathrm{l}}\\
\vec{r}_{c} & = \frac{\vec{r}_{\mathrm{u}} + \vec{r}_{\mathrm{l}}}{2}, \vec{\Delta r} = \vec{r}_{\mathrm{u}} - \vec{r}_{\mathrm{l}}\\
\end{aligned} 
\end{equation}  

The phase shift expression should be then integrated over the detection plane to obtain the full signal. The integration leaves unaffected the term proportional to the wave-packet separation $\vec{\Delta r}$, while the contribution of the second term depends on the difference of momenta $|\vec{\Delta p}|$ and the dimension of the detection region $d$. Assuming the mean position $\vec{r}_{c}$ at the center of detection region, we can define the critical condition when this phase contribution changes the sign and thus start to rapidly vanish due to the averaging: $(|\vec{\Delta p}|/\hbar)\cdot(d/2)=\pi/2$. In our case, $|\vec{\Delta p}|/\hbar = 2\epsilon k_{\mathrm{eff}}$ and $d=30$~mm, which gives a critical value of $\epsilon_{\rm{crit}}=3.8\cdot 10^{-6}$. In the region of $\epsilon\sim\epsilon_{\rm{crit}}$ this contribution might cause some varying phase shift bias. Understanding these variations requires further modeling that is outside the scope of the present work. This bias is, however, suppressed by at least an order of magnitude for the region $\epsilon\simeq 10~\epsilon_{\rm{crit}}=0.4\cdot 10^{-4}$ that covers about 80\% of the probed $\epsilon$-span. 
%Moreover, we expect an additional suppression due to the averaging over the detected spatial cloud profile (when summing up signals from all wavepackets). 
We therefore neglect this contribution and obtain:
%and consider only the first term of the the separation phase:
\begin{equation}\label{eq:A1_parasitic_sep_phase_final}
\begin{aligned}
\Delta\Phi_{\rm{sep}} & = -\frac{m\vec{v}_{0}}{\hbar}(\vec{r}_{\mathrm{u}}(t_{4})-\vec{r}_{\mathrm{l}}(t_{4})+2\epsilon\frac{\hbar\vec{k}_{\mathrm{eff}}}{m}\Delta t_{\mathrm{det}}) = -\vec{k}_{\mathrm{eff}}\vec{v}_{0}(\Delta T_{3}+2\epsilon(T+\Delta t_{\mathrm{det}}))\\
\end{aligned} 
\end{equation} 
Note, that this expression is  identical to the first term of $\Delta\Phi_{\rm{prop}}$ (Eq.~\eqref{eq:A1_parasitic_free_phase_final}) but with an opposite sign, as one may expect for a case of Lagrangian being quadratic in position and momentum \cite{Antoine2003qto}. In particular, the dependence in the timing between the final beam-splitter pulse and the detection, $\Delta t_{\mathrm{det}}$, drops out when summing the two contributions.

We now combine all the results obtained above and express the full phase shift of the bottom spurious interferometer:
\begin{equation}\label{eq:A_parasitic_full_phase_bottom}
\begin{aligned}
%\Delta\Phi^{\rm{(B)}} & = \Delta\Phi_{\rm{r}}(r_{0}) + \Delta\Phi_{\rm{v}}(v_{0}) + \Delta\Phi_{\rm{g}+\alpha} +\frac{\hbar\keff^2}{2m}(T\epsilon-\Delta T_{3})\\
\Delta\Phi^{\rm{(B)}} & = \Delta\Phi_{\rm{r}}(r_{0}) + \Delta\Phi_{\rm{v}}(v_{0}) + \Delta\Phi_{\rm{g}+\alpha} +\frac{\hbar\keff^2}{2m}(T\epsilon-\Delta T_{3}) \equiv  \Delta\Phi_{\rm{r}_{0}} + \Delta\Phi_{\rm{v}_{0}} + \Delta\Phi^{\prime} -\frac{\hbar\keff^2}{2m}T\epsilon\\
\end{aligned} 
\end{equation}
with $\Delta\Phi^{\prime} = \Delta\Phi_{\rm{g}+\alpha} + \frac{\hbar\keff^2}{2m}\left(2T\epsilon-\Delta T_{3}\right)$ being the mean phase shift independent of the initial atomic position and velocity. A fully identical calculation for the top spurious loop retrieves the same dephasing in all but recoil parts:
\begin{equation}\label{eq:A_parasitic_full_phase_top}
\begin{aligned}
%\Delta\Phi^{\rm{(T)}} & = \Delta\Phi_{\rm{r}}(r_{0}) + \Delta\Phi_{\rm{v}}(v_{0}) + \Delta\Phi_{\rm{g}+\alpha}+\frac{\hbar\keff^2}{2m}(3T\epsilon-\Delta T_{3})\\
\Delta\Phi^{\rm{(T)}} & = \Delta\Phi_{\rm{r}}(r_{0}) + \Delta\Phi_{\rm{v}}(v_{0}) + \Delta\Phi_{\rm{g}+\alpha}+\frac{\hbar\keff^2}{2m}(3T\epsilon-\Delta T_{3}) \equiv  \Delta\Phi_{\rm{r}_{0}} + \Delta\Phi_{\rm{v}_{0}} + \Delta\Phi^{\prime} +\frac{\hbar\keff^2}{2m}T\epsilon\\
\end{aligned}
\end{equation}
The total phase shifts of the two spurious loops are thus slightly different, such that $\Delta\Phi^{\rm{(T)}}-\Delta\Phi^{\rm{(B)}}=\frac{\hbar\keff^2}{m}T\epsilon$. This difference arises from the recoil terms in the free propagation contribution and vanishes for $\epsilon \to 0$. 

%Another interesting observation is that for the interferometric sequence with collinear outputs ($\epsilon=0$) one has: $\Delta\Phi_{\rm{prop}} + \Delta\Phi_{\rm{sep}} = \frac{\hbar\keff^2}{2m}(\Delta T_{3}-2\Delta T_{\rm{a}})$, which equans zero only if the time symmetry of the sequence is restored for $\Delta T_{3}= 2\Delta T_{\rm{a}}$.    

\section*{S2. Contrast of the spurious loops}
\label{sec:AppendixB}

The employed fluorescence detection in our apparatus does not discriminate the signals coming from two spurious loops. 
%Thus, in the absence of the cross-interference between the branches belonging to different loops, the detected peak-peak contrast is given by the peak-peak variation of the sum of the two spurious loops' contrasts. 
The total peak-peak contrast is therefore given by an incoherent sum of the two spurious signals. Considering the wave-packet with initial classical velocity $\vec{v}_{0}$ and position $\vec{r}_{0}$ we write:
\begin{equation}\label{eq:B_parasitic_contract_general}
C = \left[\frac{C^{\rm{(B)}}}{2}\cos{\Delta\Phi^{\rm{(B)}}}+\frac{C^{\rm{(T)}}}{2}\cos\Delta\Phi^{\rm{(T)}}\right]_{\mathrm{pp}},\\
\end{equation} 
with phase shifts $\Delta\Phi^{\rm{(B)}}$ and $\Delta\Phi^{\rm{(T)}}$ defined in Equations~\ref{eq:A_parasitic_full_phase_bottom}, \ref{eq:A_parasitic_full_phase_top} and $[...]_{\rm{pp}}$ denoting the peak-peak variation, and $C^{\rm{(B)}}$ ($C^{\rm{(T)}}$) being the contrasts of the bottom (top) spurious interferometer. We introduce the mean contrast $C_{0}=\frac{(C^{\rm{(B)}}+C^{\rm{(T)}})}{2}$, the contrasts imbalance $\Delta C_{0} = C^{\rm{(T)}}-C^{\rm{(B)}}$, the mean dephasing $\overbar{\Delta\Phi} = \frac{\Delta\Phi^{\rm{(T)}}+\Delta\Phi^{\rm{(B)}}}{2}$, and recall that $\Delta\Phi^{\rm{(T)}}-\Delta\Phi^{\rm{(B)}}=\frac{\hbar\keff^{2}}{m}T\epsilon$. The Equation~\ref{eq:B_parasitic_contract_general} becomes: 

\begin{equation}\label{eq:B_parasitic_contract_imbalanced}
\begin{aligned}
C & = C_{0}\bigg[\cos\overbar{\Delta\Phi}\cos\left(\frac{\hbar\keff^{2}}{2m}T\epsilon\right) - \frac{\Delta C_{0}}{C_{0}}\sin\overbar{\Delta\Phi}\sin\left(\frac{\hbar\keff^{2}}{2m}T\epsilon\right)\bigg]_{\mathrm{pp}}\\
\end{aligned}
\end{equation}

\begin{figure}[!h]
	\centering
	\includegraphics[width=0.48\textwidth]{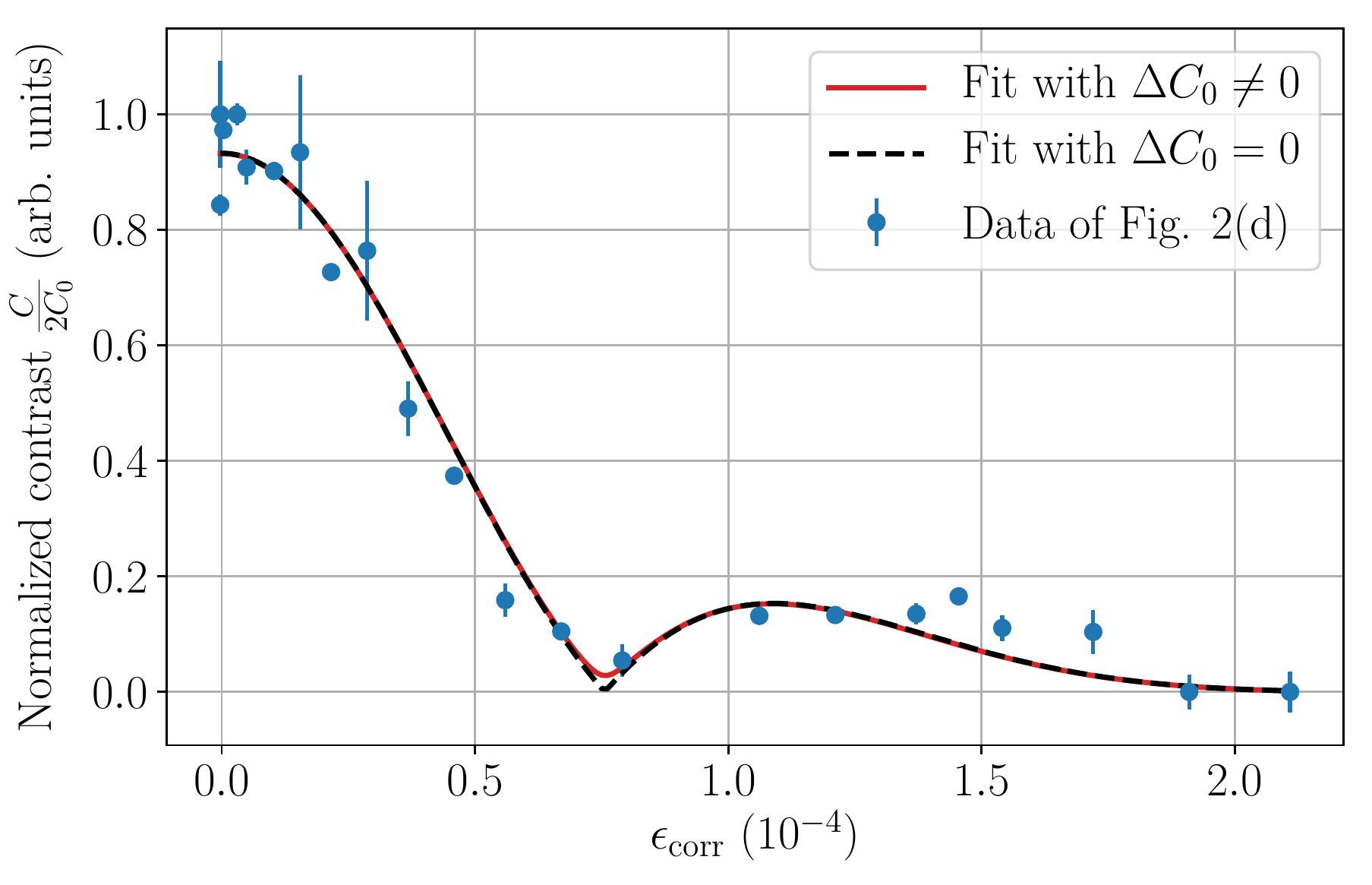}
	\caption{Normalized contrast of the spurious interferometers (data of Fig. 2(d), blue dots), fitted with exact model of Eqn.~\ref*{eq:B_parasitic_contrast_averaged} (solid red line) and simplified model (Eqn.~\ref*{eq:B_parasitic_contrast_averaged} with $\Delta C_{0}=0$, dashed black line), for comparison.} 
	\label{fig:B_parasitic_contrast_calcul}
\end{figure}

The observed contrasts of both spurious loops results from the averaging over the same initial velocity and position distributions in the atomic cloud. Assuming fully uncorrelated normal velocity ($\propto e^{-v_{0}^2/2\sigma_{v}^2}$) and position ($\propto e^{-r_{0}^2/2\sigma_{r}^2}$) distributions, we average the velocity- and position-dependent parts of the mean phase $\overbar{\Delta\Phi}$ in Eq.~\ref*{eq:B_parasitic_contract_imbalanced} and obtain normalized full peak-peak contrast as:  
%\begin{equation}\label{eq:parasitic_contrast_averaged}
%\begin{aligned}
%& \frac{C(\epsilon, \Delta T_{3})}{2C_{0}} = \bigg[\cos \Delta\tilde{\Phi}\cos\left(\epsilon\frac{\hbar\keff^{2}}{2m}T\right)-\\
%& - \frac{\Delta C_{0}}{2C_{0}}\sin \Delta\tilde{\Phi}\sin\left(\epsilon\frac{\hbar\keff^{2}}{2m}T\right)\bigg]_{\mathrm{p-p}}\times\\
%& \times\exp{\left(-\frac{\keff^{2}}{2}\left(\sigma_{v}(2\epsilon(T+t_{1})+2\Delta T_{\rm{a}}-\Delta T_{3})^{2} +(2\sigma_{r}\epsilon)^{2}\right)\right)},\\
%%& \Delta\tilde{\Phi} = \Delta\Phi_{\rm{g}+\alpha} + \frac{\hbar\keff^2}{2m}\left(2\Delta T_{\rm{a}}-\Delta T_{3} + 2T\epsilon\right)\\
%\end{aligned}
%\end{equation}
\begin{equation}\label{eq:B_parasitic_contrast_averaged}
\begin{aligned}
& \frac{C(\epsilon, \Delta T_{3})}{2C_{0}} = \exp{\left(-\frac{(2\keff\sigma_{r}\epsilon)^{2}}{2}\right)}\times\exp{\left(-\frac{(\keff\sigma_{v}(2\epsilon(T+t_{1})-\Delta T_{3}))^{2}}{2}\right)}\times\\ 
& \times\frac{1}{2}\bigg[\cos\Delta\Phi^{\prime}\cos\left(\frac{\hbar\keff^{2}}{2m}T\epsilon\right)-\frac{\Delta C_{0}}{C_{0}}\sin\Delta\Phi^{\prime}\sin\left(\frac{\hbar\keff^{2}}{2m}T\epsilon\right)\bigg]_{\mathrm{pp}}\\
%& \Delta\tilde{\Phi} = \Delta\Phi_{\rm{g}+\alpha} + \frac{\hbar\keff^2}{2m}\left(2\Delta T_{\rm{a}}-\Delta T_{3} + 2T\epsilon\right)\\
\end{aligned}
\end{equation}
%with $\Delta\Phi^{\prime} = \Delta\Phi_{\rm{g}+\alpha} + \frac{\hbar\keff^2}{2m}\left(2T\epsilon-\Delta T_{3}\right)$ being the mean phase shift independent of the statistical distribution in the atomic cloud.
%The exponential terms represent the suppression of the contrast due to the finite size and momentum thermal width of the atomic cloud, causing spatial and momentum decoherence. The last term reveals oscillations with a period being independent on the initial velocity and position statistical distributions. The first zero of the contrast is thus expected around the value of $\epsilon=0.76\cdot 10^{-4}$ the interference patterns of the two spurious loops are in anti-phase.
%In the case of equal contrasts ($\Delta C = 0$), we have simply: 
%\begin{equation}\label{eq:B_parasitic_contract_simple}
%C_{\mathrm{p-p}} = 2C\bigg|\cos\left(\epsilon \frac{\hbar\keff^2}{2m}T\right)\bigg|\\
%\end{equation} 
%The observed zero-point matches this value within about 10\% accuracy. 
In Figure~\ref{fig:B_parasitic_contrast_calcul} we show the fit of the data with the general-case model of Eqn.~\ref*{eq:B_parasitic_contrast_averaged}, where mean peak contrast $C_{0}$, contrast imbalance $\Delta C_{0}$ and standard deviation $\sigma_{r}$ are free parameters (solid red line). We extract the values of $\sigma_{r}=0.51(2)$~mm, $2C_{0}=0.934(18)$  and $\Delta C_{0}=0.03(3)\%$. The value of $2C_{0}<1$ simply accounts for the actual over-estimation of the maximum contrast resulting from data normalization to the maximum of the recorded contrasts. The fitted contrast imbalance $\Delta C_{0}=0.03(3)\%$ is well compatible with zero. Comparing this fit with the fit by simplified model used in the main text (dashed black line, for $\Delta C_{0}=0$) shows a small difference around the contrast local minimum at $\epsilon=0.76\cdot 10^{-4}$, without any change for the rest of the probed $\epsilon$-values. Thus, all the arguments presented in the main text remain true for the case of the fit with exact function accounting for small contrast imbalance of the two spurious interferometers.
%The overall trend is reproduced by the curves corresponding to up to 20\% imbalance. The largest discrepancy is observed around the revival peak, whose amplitude is about factor of 2 lower than expected. One possible explanation may be that the envelope curve in reality features more complex dependence on $\epsilon$ due to the deviation of the initial cloud shape from the Gaussian function. 
%and features a stronger decay for the values of $\epsilon\gtrsim 0.7\cdot 10^{-4}$. 
%On the other hand, we may fit the data with the $\sigma_{r}$ and amplitude parameters free while keeping the contrast imbalance at zero. The result is shown in Figure~\ref{fig:B_parasitic_contrast_fit}. 
%\begin{figure}[t]
%	\centering
%	\includegraphics[width=0.48\textwidth]{Figures/FigB2.pdf}
%	\caption{(a) Comparison of the two fitting functions for the peak-peak contrast data.} 
%	\label{fig:B_parasitic_contrast_fit}
%\end{figure} 

%\input acknowledgement.tex   % input acknowledgement

\section*{S3. Additional data on spurious interferometers}
\label{sec:AppendixC}

\begin{figure}[!h]
	\centering
	\includegraphics[width=0.48\textwidth]{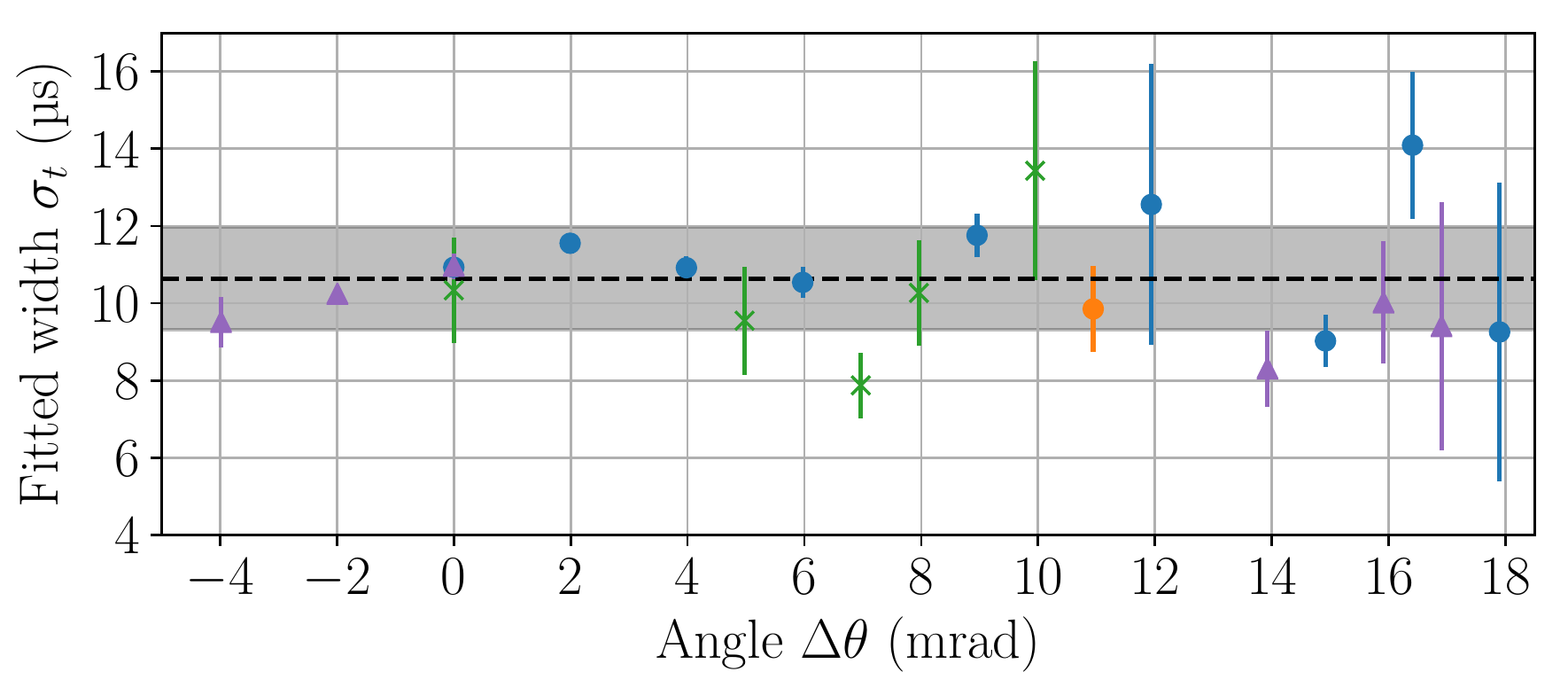}
	\caption{Fitted time-domain width of the spurious peaks for the probed values of $\Delta\theta$. Various symbols and accompanying colors indicate data sets taken on different days within two-week period. The dashed black line (gray-shaded area) are the weighted mean (standard deviation interval) of all shown data.}
	\label{fig:C_spurious_sigma_t}
\end{figure}

\begin{figure}[!h]
	\centering
	\includegraphics[width=0.65\textwidth]{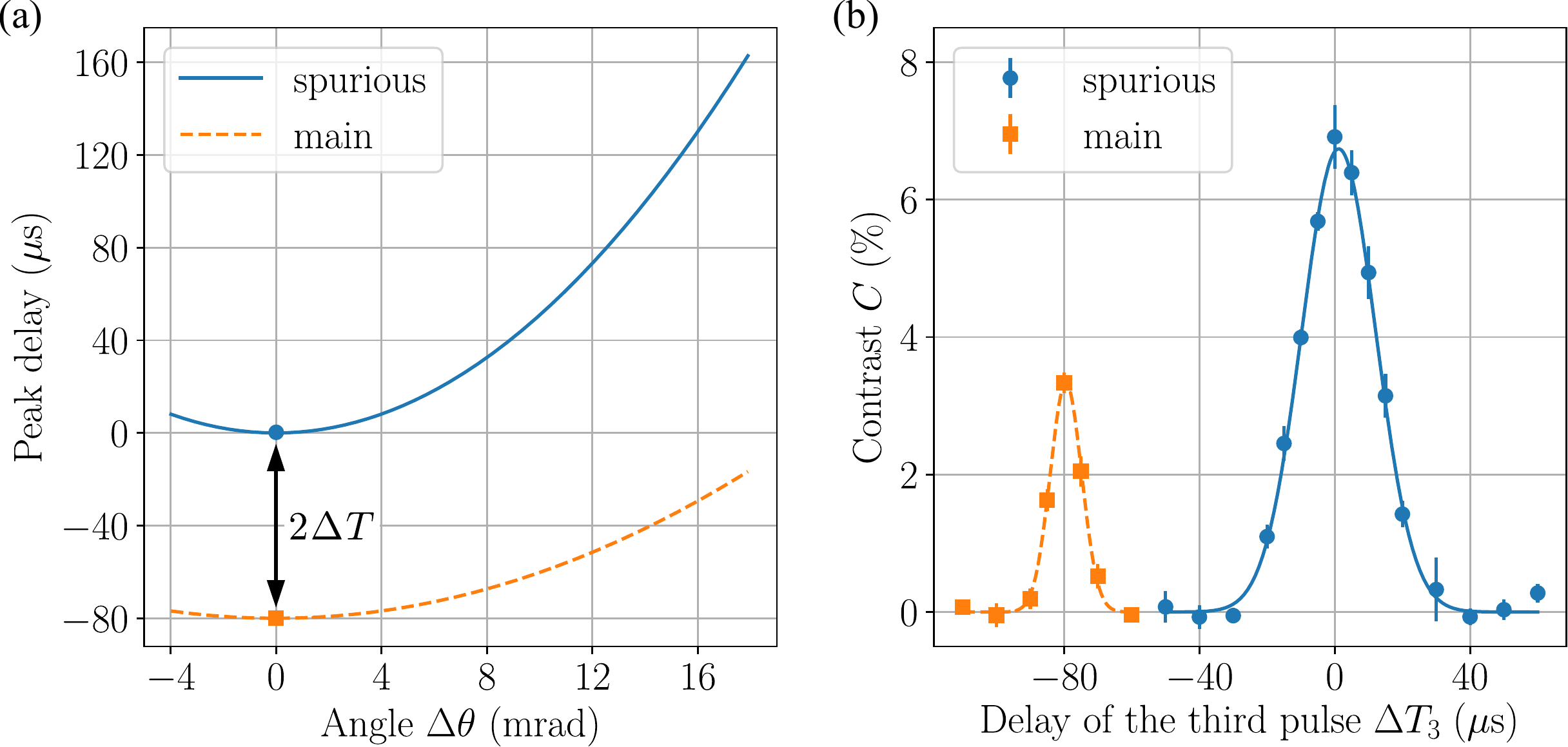}
	\caption{(a) Expected peak delay for spurious (solid blue line) and main (dashed orange line) interferometers as a function of $\Delta\theta$, for initial time separation $\Delta T=40~\mu\rm{s}$. The blue dot and orange square mark the expected peak positions for the data shown in the panel (b). (b) Peak-peak contrast of the spurious (blue dots) and main (orange squares) as a function of the third laser pulse delay $\Delta T_{3}$, for $\Delta\theta=0$, and Gaussian fits (solid blue and dashed orange lines) to the corresponding data.}
	\label{fig:C_spurious_and_main}
\end{figure}

\noindent\textbf{\textit{Time-domain width}} 
In Figure~\ref{fig:C_spurious_sigma_t} we show the extracted the time-domain widths of the spurious interferometric peaks $\sigma_{t}$ for all data sets similar to those of the Figure~2(b). The data shows a rather large scatter for the probed range of $\Delta\theta$ which is likely to come from an hour-timescale experimental variations, and day-to-day drifts in case of different data sets. In overall, we cannot identify any clear systematic trend and the behavior seems consistent with the expected independence of $\Delta\theta$. We thus obtain a weighted mean value of $\bar{\sigma}_{t}=10.6(1.3)~\mu\rm{s}$ (dashed black line in Fig.~\ref{fig:C_spurious_sigma_t}) that we use to empirically set the value of $\sigma_{v}=1/\keff\bar{\sigma}_{t}=1.8(2)~v_{R}$, where $v_{R}$ is the single-photon atom recoil velocity. This value differs from the initial thermal width of $3.0(2)~v_{R}$, underlining the impact of the velocity-selection during the interrogation pulses.
\\
\\
\noindent\textbf{\textit{Time-separation of the spurious and main interferometers}}
While studying controlled recombination of the spurious interferometers, it is important to verify that the main interferometer is sufficiently distant such that its wings do not affect the spurious signal. In Figure~\ref{fig:C_spurious_and_main}(a) we plot the expected peak recombination time moment for spurious (same as solid black line in Fig.~2(c) of the main text) and main interferometers. These functions are given by: $\Delta T_{3}=(T+t_{1})\Delta\theta^{2}$ (spurious interferometer, solid blue line) and $\Delta T_{3}=-2\Delta T + \frac{T}{2}\Delta\theta^{2}$ (main interferometer, dashed orange line), with $\Delta T=40~\mu\rm{s}$ being an initial time shift of the second and third pulses as explained in the main paper. The timing separation between two peaks, therefore, is minimal and equals $2\Delta T$ for $\Delta\theta=0$ and increases with increasing $|\Delta\theta|$. In Figure~\ref{fig:C_spurious_and_main}(b) we demonstrate that the choice of $\Delta T=40~\mu\rm{s}$ excludes any overlap between the two peaks for $\Delta\theta=0$.       

\section*{S4. Phase shift of the main loop}
\label{sec:AppendixD}

\noindent\textbf{\textit{Sensitivity to rotation rate}} We derive the sensitivity to rotation rate of the main double-loop interferometer for the perfectly recombined symmetric configuration considered in the main paper using three different methods: the ABCD-matrix formalism \cite{Antoine2003qto}, the full phase shift calculation approach (similar to the one of spurious intrferometers), and the geometric approach of Sagnac area calculation. All methods give the same result:
\begin{equation}\label{eq:D_rotation_phase_shift}
\begin{aligned}
\Delta\Phi_{\Omega} =\frac{1}{2}\vec{k}_{\rm{eff}}(\vec{g}\times\vec{\Omega})T^{3}\left(1-\frac{2\epsilon}{3}\right)\\
\end{aligned} 
\end{equation} 
%\red{Should we give a drawing and a simple derivation of the geometrical area correction?}\\ 
%Considering the rotation rate of the Earth at our latitude (Paris), $\Omega=4.8\cdot 10^{-5}$~rad/s, the correction to the gyroscope scale factor would produce (of order $10^{-4}$~rad) could not be resolved in this study.}
%The first term (called hereafter clock shift) represents the sensitivity to the detuning ($\Delta\omega_0$) of the relative Raman  laser frequencies from the  resonance condition of the Raman transition at the apogee point  (see Supplemental Material \cite{SupMat}). The second term accounts for a correction to the gyroscope scale factor, as can be  derived from the reduction of the physical (Sagnac) area of the interferometer. We measured the clock shift and confirmed the first term of Eq.\eqref{eq:main_phase_shift} with an accuracy of XX. Such a clock shift is conventionally suppressed by alternating measurements with $\pm \keff$, allowing us to achieve a suppression by a factor YY. The correction to the gyroscope scale factor (of order $10^{-4}$) could not be resolved in this study.
\\
\noindent\textbf{\textit{Sensitivity to frequency}} An additional phase shift may arise in the AMT configuration if the effective laser frequency is detuned from the resonance condition at the apogee point of the fountain trajectory by a fixed amount $\Delta\omega_{0}$. This so-called clock shift can be estimated with by accounting for the frequency contribution to the imprinted laser phase (similarly to the above calculation for the spurious interferometers), or via sensitivity function \cite{Cheinet2008} approach. In the limit of infinitely short laser pulses, we obtain:   
\begin{equation}\label{eq:D_clock_phase_shift}
\begin{aligned}
\Delta\Phi_{\rm{clock}} = 4\Delta\omega_{0}\Delta T_{\rm{s}} = \Delta\omega_{0}\frac{2T\epsilon}{(1-\epsilon)}\approx\Delta\omega_{0}T\Delta\theta^{2}\\
\end{aligned} 
\end{equation}  

\begin{figure}[!h]
	\centering
	\includegraphics[width=0.53\textwidth]{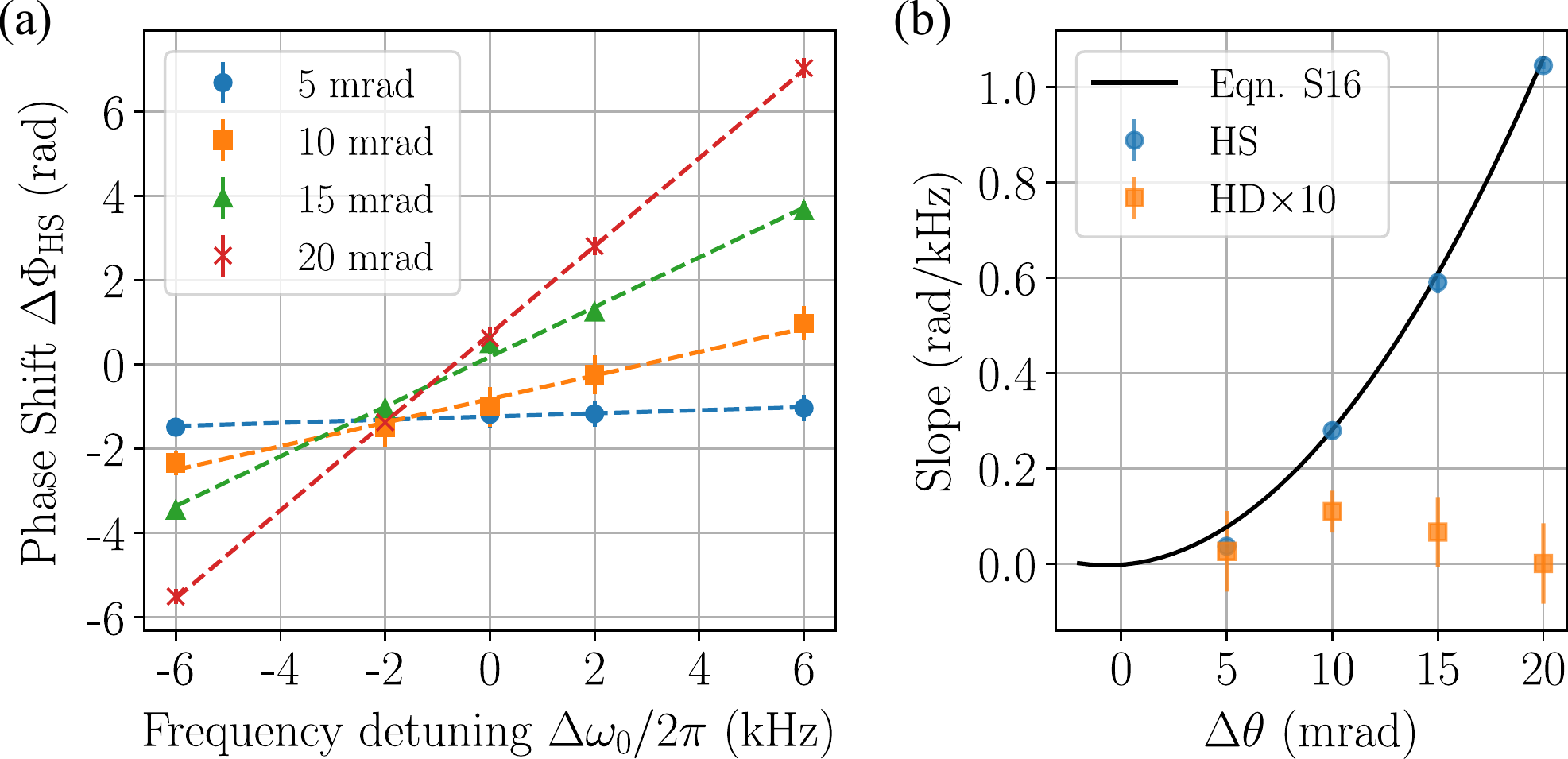}
	\caption{(a) Sensitivity of the main interferometer to the two-photon frequency in the AMT scheme, for different values of $\Delta\theta$, as extracted from the half-sum (HS) of the $\pm\keff$ measurements. The solid blue, dashed orange, dash-dotted green and dotted red curves are linear fits to the corresponding data. (b) The fitted slopes $d\Delta\Phi/d(\Delta\omega_{0}/2\pi)$ of the half-sum signal of panel (a) and half-difference signal (HD), as a function of probed $\Delta\theta$. The solid black line indicates the expectation given by the Eqn.~\ref{eq:D_clock_phase_shift}  for $\epsilon=\epsilon_{\rm{calc}}$. The HD data and error bars are increased by a factor of 10 for visibility.}
	\label{fig:main_clock_shift}
\end{figure}

To quantify the clock sensitivity, we record the induced phase shift from the controlled change of the two-photon detuning  for a set of different angles. The phase shift $\Delta\Phi_{\rm{HS}}$ is evaluated as a half-sum (HS) of the measured values for alternating sign of $\vkeff$ and shows the expected linear dependence on frequency detuning $\Delta\omega_{0}$ (Fig.~\ref{fig:main_clock_shift}(a)). The fitted linear slopes $d\Delta\Phi_{\rm{HS}}/d(\Delta\omega_{0}/2\pi)$ scale quadratically with $\Delta\theta$ (blue dots in Fig.~\ref{fig:main_clock_shift}(b)), well matched with the expectation from Eq.~\ref{eq:D_clock_phase_shift} (solid black line). As the clock shift is independent on $\vkeff$, it should vanish (or be significantly suppressed) in the half-difference (HD) signal of $\pm\keff$ method that leaves unaffected the inertial shifts. The orange squares in Figure ~\ref{fig:main_clock_shift}(b) show the corresponding clock sensitivity given by $d\Delta\Phi_{\rm{HD}}/d(\Delta\omega_{0}/2\pi)$, boosted by a factor of 10 (including the error bars) for better visibility. We estimate a suppression factor ranging from about 10 (at 5~mrad) to better than 100 (at 20~mrad).

%From the mean value of the slope 5(8)~mrad/kHz,

%We further extract the time-domain widths of the echo peaks $\sigma_{t}$ (Fig.~\ref{fig:parasitic_interferometers}(c)), that is directly related to $\sigma_{v}$ via equation~\ref{eq:parasitic_contrast}. The data shows a rather large scatter for the probed range of $\Delta\theta$ which is likely to come from an hour-timescale experimental variations, and day-to-day drifts in case of different data sets. In overall, we cannot identify any clear systematic trend and the behavior seems consistent with the expected independence of $\Delta\theta$. We thus fit the data with a constant and extract the effective value of $\sigma_{v}=1.81()~v_{R}$, where $v_{R}$ is the atom recoil velocity. This value differs from the initial thermal width of $3~v_{R}$, underlining the impact of the velocity-selection during the interrogation pulses.


\end{document}